\documentstyle[12pt]{article}
\begin{document}


\renewcommand{\thesection}{\arabic{section}}
\renewcommand{\theequation}{\arabic{equation}}
\renewcommand {\c}  {\'{c}}
\newcommand {\cc} {\v{c}}
\newcommand {\s}  {\v{s}}
\newcommand {\CC} {\v{C}}
\newcommand {\C}  {\'{C}}
\newcommand {\Z}  {\v{Z}}
\newcommand{\pv}[1]{{-  \hspace {-4.0mm} #1}}

\baselineskip=24pt

\def\beqra{\begin{eqnarray}} \def\eeqra{\end{eqnarray}}
\def\beqast{\begin{eqnarray*}} \def\eeqast{\end{eqnarray*}}
\def\beq{\begin{equation}}      \def\eeq{\end{equation}}
\def\be{\begin{enumerate}}   \def\ee{\end{enumerate}}

\def\gam{\gamma}
\def\Gam{\Gamma}
\def\la{\lambda}
\def\eps{\epsilon}
\def\La{\Lambda}
\def\si{{\rm si}}
\def\Si{\Sigma}
\def\al{\alpha}
\def\Th{\Theta}
\def\th{\theta}
\def\tnu{\tilde\nu}
\def\vphi{\varphi}
\def\del{\delta}
\def\Del{\Delta}
\def\ab{\alpha\beta}
\def\om{\omega}
\def\Om{\Omega}
\def\mn{\mu\nu}
\def\mun{^{\mu}{}_{\nu}}
\def\kap{\kappa}
\def\rsi{\rho\sigma}
\def\beal{\beta\alpha}
\def\til{\tilde}
\def\rta{\rightarrow}
\def\eqv{\equiv}
\def\nab{\nabla}
\def\pa{\partial}
\def\sit{\tilde\sigma}
\def\ul{\underline}
\def\indt{\parindent2.5em}
\def\nd{\noindent}
\def\rsi{\rho\sigma}
\def\beal{\beta\alpha}
\def\caa{{\cal A}}
\def\cb{{\cal B}}
\def\cac{{\cal C}}
\def\cd{{\cal D}}
\def\ce{{\cal E}}
\def\cf{{\cal F}}
\def\cg{{\cal G}}
\def\cah{{\cal H}}
\def\ci{{\rm ci}}
\def\cj{{\cal{J}}}
\def\ck{{\cal K}}
\def\cl{{\cal L}}
\def\cm{{\cal M}}
\def\cn{{\cal N}}
\def\cO{{\cal O}}
\def\cp{{\cal P}}
\def\car{{\cal R}}
\def\cs{{\cal S}}
\def\ct{{\cal{T}}}
\def\cu{{\cal{U}}}
\def\cv{{\cal{V}}}
\def\cw{{\cal{W}}}
\def\cx{{\cal{X}}}
\def\cy{{\cal{Y}}}
\def\cz{{\cal{Z}}}
\def\asymptotic{{_{\stackrel{\displaystyle\longrightarrow}
{x\rightarrow\pm\infty}}\,\, }} 
\def\asymptext{\raisebox{.6ex}{${_{\stackrel{\displaystyle\longrightarrow}
{x\rightarrow\pm\infty}}\,\, }$}} 
\def\epsilim{{_{\textstyle{\rm lim}}\atop
_{~~~\epsilon\rightarrow 0+}\,\, }} 
\def\omegalim{{_{\textstyle{\rm lim}}\atop
_{~~~\om^2\rightarrow 0+}\,\, }} 
\def\xlimp{{_{\textstyle{\rm lim}}\atop
_{~~x\rightarrow \infty}\,\, }} 
\def\xlimm{{_{\textstyle{\rm lim}}\atop
_{~~~x\rightarrow -\infty}\,\, }} 
\def\asymptoticp{{_{\stackrel{\displaystyle\longrightarrow}
{x\rightarrow +\infty}}\,\, }} 
\def\asymptoticm{{_{\stackrel{\displaystyle\longrightarrow}
{x\rightarrow -\infty}}\,\, }} 

\def\raisenot{\raise .5mm\hbox{/}}
\def\nota{\ \hbox{{$a$}\kern-.49em\hbox{/}}}
\def\notA{\hbox{{$A$}\kern-.54em\hbox{\raisenot}}}
\def\notb{\ \hbox{{$b$}\kern-.47em\hbox{/}}}
\def\notB{\ \hbox{{$B$}\kern-.60em\hbox{\raisenot}}}
\def\notc{\ \hbox{{$c$}\kern-.45em\hbox{/}}}
\def\notd{\ \hbox{{$d$}\kern-.53em\hbox{/}}}
\def\notbd{\ \hbox{{$D$}\kern-.61em\hbox{\raisenot}}} 
\def\note{\ \hbox{{$e$}\kern-.47em\hbox{/}}}
\def\notk{\ \hbox{{$k$}\kern-.51em\hbox{/}}}
\def\notp{\ \hbox{{$p$}\kern-.43em\hbox{/}}}
\def\notq{\ \hbox{{$q$}\kern-.47em\hbox{/}}}
\def\notW{\ \hbox{{$W$}\kern-.75em\hbox{\raisenot}}}
\def\notz{\ \hbox{{$Z$}\kern-.61em\hbox{\raisenot}}}
\def\notpa{\hbox{{$\partial$}\kern-.54em\hbox{\raisenot}}}
\def\fo{\hbox{{1}\kern-.25em\hbox{l}}}  
\def\rf#1{$^{#1}$}
\def\bx{\Box}
\def\tr{{\rm Tr}}
\def\rmtr{{\rm tr}}
\def\dgg{\dagger}
\def\lag{\langle}
\def\rag{\rangle}
\def\bmid{\big|}
\def\vlap{\overrightarrow{\La p}} 
\def\lrta{\longrightarrow} \def\lrar{\raisebox{.8ex}{$\longrightarrow$}}
\def\ON{{\cal O}(N)}
\def\UN{{\cal U}(N)}
\def\bdPh{\mbox{\boldmath{$\dot{\!\Phi}$}}}
\def\bPh{\mbox{\boldmath{$\Phi$}}}
\def\bPhs{\bPh^2}
\def\sef{S_{eff}[\sigma,\pi]}
\def\sigx{\sigma(x)}
\def\pix{\pi(x)}
\def\bph{\mbox{\boldmath{$\phi$}}}
\def\bphs{\bph^2}
\def\ex{\BM{x}}
\def\exs{\ex^2}
\def\xdot{\dot{\!\ex}}
\def\y{\BM{y}}
\def\ys{\y^2}
\def\ydot{\dot{\!\y}}
\def\pat{\pa_t}
\def\pax{\pa_x}
\def\hp{{\pi\over 2}}
\def\sign{{\rm sign}\,}
\def\bv{{\bf v}}




\begin{center}
{\bf  Density Waves in the Calogero Model - Revisited}
\bigskip

V. Bardek$^{a}${ \footnote{e-mail: bardek@irb.hr}},
J. Feinberg$^{b,c,d}${ \footnote{e-mail: joshua@physics.technion.ac.il}}
 \hspace{0.2cm} and \hspace{0.2cm}
S.Meljanac$^{a}$ {\footnote{e-mail: meljanac@irb.hr}} \\
$^{a}$ Rudjer Bo\v{s}kovi\'c Institute, Bijeni\v cka  c.54, HR-10002 Zagreb,
Croatia \\[3mm]

$^{b}$  Department of Physics, University of Haifa at Oranim, Tivon 36006,
Israel, \\

$^{c}$ Department of Physics, Technion-Israel Inst. of Technology,
Haifa 32000, Israel,\newline and \\

$^{d}$ KITP, University of California, Santa Barbara, CA 93106-4030, USA \\[3mm]

\bigskip

\end{center}
\setcounter{page}{1}
\bigskip


\begin{minipage}{5.8in}

{\abstract~~~The Calogero model bears, in the continuum limit,
collective excitations in the form of density waves and solitary
modulations of the density of particles. This sector of the
spectrum of the model was investigated, mostly within the
framework of collective field theory, by several authors, over the
past fifteen years or so. In this work we shall concentrate on
periodic solutions of the collective BPS-equation (also known as
``finite amplitude density waves"), as well as on periodic solutions of
the full static variational equations which vanish periodically
(also known as ``large amplitude density waves"). While
these solutions are not new, we feel that our analysis and
presentation add to the existing literature, as we explain in the
text. In addition, we show that these solutions also occur in a
certain two-family generalization of the Calogero model, at
special points in parameter space. A compendium of useful identities associated with
Hilbert transforms, including our own proofs of these identities,
appears in Appendix A. In Appendix B we also elucidate in
the present paper some fine points having to do with manipulating
Hilbert transforms, which appear ubiquitously in the collective
field formalism. Finally, in order to make this paper self-contained, we briefly 
summarize in Appendix C basic facts about the collective field formulation of the Calogero model.}

\end{minipage}

\bigskip
PACS number(s): 03.65.Sq, 05.45.Yv, 11.10.Kk, 11.15.Pg \\
\bigskip
\bigskip
Keywords: Calogero model, collective-field theory, BPS, solitons


\newpage


\section{Introduction}
\setcounter{footnote}{0}
The Calogero Model (CM) \cite{Calogero:1969xj} -
\cite{Olshanetsky:1981dk} is a well-known exactly solvable
many-body system, both at the classical and quantum levels. It
describes $N$ particles (considered indistinguishable at the
quantum level) on the line, which interact through an
inverse-square two-body interaction. Its quantum Hamiltonian is
\begin{equation} \label{Hcalogero}
H = - \frac{1}{2 m} \sum_{i=1}^{N} \frac{{\partial}^{2}}{\partial
{x_{i}}^{2}} + \frac{\lambda (\lambda - 1)}{2 m} \sum_{i \neq j
}^{N} \frac{1}{{(x_{i} - x_{j})}^{2}}\,,
\end{equation}
where $m$ is the particles' mass, and the dimensionless coupling
constant $\lambda$ parametrizes the inverse-square interaction
between pairs of particles. \footnote{Note that we did not include
in (\ref{Hcalogero}) a confining potential. This is not really a
problem, as we can always add a very shallow confining potential
to regulate the problem (in the case of purely repulsive
interactions), or else, consider the particles confined to a very
large circle (i.e., consider (\ref{Hcalogero}) as the large radius
limit of the Calogero-Sutherland model \cite{Sutherland:1971ep}).
We shall henceforth tacitly assume that the system is thus
properly regularized at large distances.}

The CM and its various descendants continue to draw considerable
interest due to their many diverse physical applications. A
partial list of these applications can be found, for example, in
the introductory section of \cite{BFM}. For recent reviews on the
Calogero- and related models see, e.g.,
\cite{Polychronakos:1999sx, Polychronakos:2006}. In addition, for
a recent review on the collective-field and other continuum
approaches to the spin-Calogero-Sutherland model, see
\cite{Aniceto-Jevicki}.

In the present paper we concentrate on the thermodynamic limit of
the CM. In this limit the system is amenable to large-$N$
collective-field formulation \cite{sakita, Jevicki:1979mb, JFnmm}.
As is well-known, the collective theory offers a continuum
field-theoretic framework for studying interesting aspects of
many-particle systems. Clearly, a description of the particle
systems in terms of continuous fields becomes an effectively good
one in the high density limit. In this limit the mean
interparticle distance is much smaller than any relevant
physical length-scale, and the $\delta$-function spikes in the
density field (\ref{collective}) below can be smoothed-out into a
well-behaved countinuum field. All this is in direct analogy to
the hydrodynamical effective description of fluids, which replaces
the microscopic atomistic formulation. Of course, the large
density limit means that we have taken the large- $N$ limit, as
was mentioned above.

The collective-field Hamiltonian for the CM (\ref{Hcalogero}) is
given by \cite{AJL} \beq\label{Hcollective} H_{coll} = \frac{1}{2
m} \int dx\, \pax\pi(x)\, \rho(x)\, \pax\pi(x) + \frac{1}{2 m}
\int dx\, \rho(x) {\left( \frac{\lambda - 1}{2} \frac{\partial_{x}
\rho}{\rho} + \lambda \pv \int \frac{ dy \rho(y)}{x - y}
\right)}^{2}  + H_{sing}\,, \eeq where  $ \; H_{sing} \; $ denotes
a singular contribution \cite{Andric:1994su}
\beq\label{Hsing} H_{sing} =  - \frac{\lambda}{2 m}\,\int dx\,
\rho(x)\,\partial_{x}\left. \frac{P}{x - y} \right|_{y = x} -
\frac{\lambda - 1}{4 m}\,\int dx\, {\partial_{x}}^{2} \left.
\delta(x - y) \right|_{y = x}\,, \eeq and $ \; P \;$ is the
principal part symbol.

Here, \beq\label{collective} \rho(x) = \sum_{i = 1}^{N} \delta( x
- x_{i}) \eeq is the collective  - or density - field, and
\beq\label{momenta} \pi(x) = - i \frac{\delta}{\delta \rho(x)}
\end{equation}
is its canonically conjugate momentum. It follows trivially from
(\ref{collective}) that the collective field is a positive
operator \beq\label{positiverho} \rho(x) \geq 0\,,\eeq and that it
obeys the normalization condition \beq\label{conservation}
\int\limits_{-\infty}^\infty\,dx \,\rho (x) = N\,. \eeq The latter
constraint is implemented by adding to (\ref{Hcollective}) a term
$\mu\left(\int\limits_{-\infty}^\infty\,dx \,\rho (x) - N\right)$,
where $\mu$ is a Lagrange multiplier (the chemical potential). 

The first term in (\ref{Hsing}) is proportional to $\rho(x)$. Therefore, its singular coefficient
$-{\lambda\over 2m}\partial_{x}\left. \frac{P}{x - y} \right|_{y = x} $ amounts to a shift of the 
chemical potential $\mu$ by an infinite constant. The last term in (\ref{Hsing}) is, of course, 
a field independent constant - an infinite shift of energy. In order for this paper to be self-contained, we 
have briefly summarized the derivation of the collective-field Hamiltonian (\ref{Hcollective}) in Appendix C.

It is worth mentioning at this point that the Calogero model
enjoys a strong-weak-coupling duality symmetry
\cite{Minahan:1994ce, halzirn}. At the level of the collective
Hamiltonian (\ref{Hcollective}), these duality transformations
read \beq\label{duality} \tilde\lambda = {1\over\lambda}\,,\quad
\tilde m = -{m\over\lambda}\,,\quad \tilde \mu =
-{\mu\over\lambda}\,;\quad \tilde\rho(x) = -\lambda\rho(x)\,,\quad
{\rm and}\quad \tilde\pi(x) = -{\pi(x)\over\lambda}\,,\eeq and it
is straightforward to see that these transformations leave
(\ref{Hcollective}) (including the chemical potential term)
invariant. The minus signs which occur in (\ref{duality}) are all
important: We interpret all negative values of the parameters and
densities as those pertaining to holes, or antiparticles. Thus,
the duality transformations (\ref{duality}) exchange particles and
antiparticles. (For more details see e.g. Section 3 of \cite{BFM},
and references therein.)

It is well-known \cite{Jevicki:1979mb} that to leading order in
the ${1\over N}$ expansion, collective dynamics of our system is
determined by the classical equations of motion resulting from
(\ref{Hcollective}). The simplest solution of these equations is
the constant condensate $\rho (x) = \rho_0$ (and $\pi (x) = 0$)
corresponding to the ground state.

More interesting solutions of
these equations include various types of periodic density waves and soliton
configurations \cite{Polychronakos:1994xg, Andric:1994nc,
Sen:1997qt}. As we explain in Section 2 below, these periodic density waves
can be thought of as a crystal made of the localized soliton solution.

Recently, density wave configurations of
this type were studied, 
among other things, in \cite{ajj2}, where
a certain regulator, first introduced in \cite{Andric:1994nc}, was
used to tame the effective collective potential. The BPS-equations
associated with the regulated potential were then converted into a
Riccati equation, which was solved explicitly.

Such static periodic density waves are the focus of the present paper
as well. As in \cite{ajj2}, we convert the BPS-equations
associated with the equations of motion of (\ref{Hcollective})
into an explicitly solvable Riccati equation. However, unlike
\cite{ajj2}, we avoid introducing any unconventional regulators in
(\ref{Hcollective}). In addition, we also construct non-BPS
solutions of the equations of motion, which are simple shifts of
BPS-solutions by a constant, and compute their energy densities.
That constant is fixed by the equation of motion and turns out to
be either the maximum or the minimum value of the corresponding
BPS-solution. Thus, these non-BPS density profiles vanish
periodically and coincide with the large-amplitude waves reported
in \cite{Sen:1997qt}, albeit without too many details of their
construction. We believe that the constructive way in which we
derive our static periodic BPS and non-BPS configurations
complements the discussion in \cite{Polychronakos:1994xg,
Sen:1997qt, ajj2}. Since these non-BPS solutions vanish
periodically, we can also refer to them as vortex crystals, as
they constitute a periodic generalization of the vortex solution
of \cite{Andric:1994nc}.

In the present paper we also show how these known solitary and
periodic wave-solutions appear in the collective field theory of the
two-family generalization of the CM, under very special conditions
on the coupling constants. The two-family Calogero model is a
generalization of (\ref{Hcalogero}) into two species of identical
particles. The Hamiltonian of this model reads
\cite{Meljanacstojic} \beqra\label{h1} H = &-& \frac{1}{2 m_{1}}
\sum_{i=1}^{N_{1}} \frac{{\partial}^{2}}{\partial {x_{i}}^{2}} +
\frac{\lambda_{1} (\lambda_{1} - 1)}{2 m_{1}} \sum_{i \neq j
}^{N_{1}} \frac{1}{{(x_{i} - x_{j})}^{2}}\nonumber\\{}\nonumber\\
&-& \frac{1}{2 m_{2}} \sum_{\alpha = 1}^{N_{2}}
\frac{{\partial}^{2}}{\partial {x_{\alpha}}^{2}} +
\frac{\lambda_{2} (\lambda_{2} - 1)}{2 m_{2}} \sum_{\alpha \neq
\beta }^{N_{2}} \frac{1}{{(x_{\alpha} -
x_{\beta})}^{2}}\nonumber\\{}\nonumber\\
&+& \frac{1}{2} \left( \frac{1}{ m_1}  +  \frac{1} { m_2 } \right)
 \lambda_{12}(\lambda_{12} -1)
 \sum_{i = 1}^{N_{1}}\sum_{\alpha = 1 }^{N_{2}}
 \frac{1}{(x_{i}-x_{\alpha})^{2}}\,.
\eeqra Here, the first family contains $ \; N_{1} \; $ particles
of mass $ \; m_{1} \; $ at positions $ \; x_{i}, \; i =
1,2,...,N_{1}, \; $ and the second one contains $ \; N_{2} \; $
particles of mass $ \; m_{2} \; $ at positions $ \; x_{\alpha}, \;
\alpha = 1,2,...,N_{2}. $ All particles interact via two-body
inverse-square potentials. The interaction strengths within each
family are parametrized by the coupling constants $ \; \lambda_{1}
\; $ and $ \; \lambda_{2}, \; $ respectively. The interaction
strength between particles of the first and the second family is
parametrized by $ \; \lambda_{12}.$

In (\ref{h1}) we imposed the restriction that there be no
three-body interactions, which requires
\cite{Meljanacstojic}-\cite{Meljsams}
\begin{equation} \label{threebody}
  \frac{\lambda_{1}}{{m_{1}}^{2}} = \frac{\lambda_{2}}{{m_{2}}^{2}} =
 \frac{\lambda_{12}}{m_{1} m_{2}}.
\end{equation}
It follows from (\ref{threebody}) that \beq\label{lambda12}
\lambda_{12}^2 = \lambda_1\lambda_2\,. \eeq We assume that
(\ref{threebody}) and (\ref{lambda12}) hold throughout this paper
wherever we discuss the two-family CM. The Hamiltonian (\ref{h1})
describes the simplest multi-species Calogero model for particles
on the line, interacting only with two-body potentials.

In \cite{BFM} we studied the collective field theory of the
two-family CM. The corresponding collective Hamiltonian is
\beqra\label{Hcollective2F} H_{coll} &=&  \frac{1}{2 m_{1}} \int
dx\, \pax\pi_1(x)\, \rho_{1}(x)\, \pax\pi_1(x)\nonumber\\{}\nonumber\\
&+& \frac{1}{2 m_{1}} \int dx \rho_{1}(x) {\left(
\frac{\lambda_{1} - 1}{2} \frac{\partial_{x} \rho_{1}}{\rho_{1}} +
\lambda_{1} \pv \int \frac{ dy \rho_{1}(y)}{x - y}  + \lambda_{12}
\pv \int \frac{dy
\rho_{2}(y)}{x - y} \right)}^{2}\nonumber\\{}\nonumber\\
 &+& \frac{1}{2 m_{2}} \int dx \,\pax\pi_2(x)\,\rho_{2}(x) \,\pax\pi_2(x)
 \nonumber\\{}\nonumber\\
 &+& \frac{1}{2 m_{2}} \int dx  \rho_{2}(x) {\left( \frac{\lambda_{2} - 1}{2}
\frac{\partial_{x} \rho_{2}}{\rho_{2}}
  + \lambda_{2} \pv \int \frac{ dy  \rho_{2}(y)}{x - y}
  + \lambda_{12} \pv \int \frac{dy \rho_{1}(y)}{x - y}
  \right)}^{2}\nonumber\\{}\nonumber\\ &+&  H_{sing}\,,\eeqra
which is a straightforward generalization of (\ref{Hcollective}).
Here $\rho_a (x)$ are the collective density fields of the $a$th
family ($a=1,2$), and $\pi_a(x)$ are their conjugate momenta. As
in (\ref{Hcollective}), the term $ \; H_{sing} \; $ denotes a
singular  contribution which is a straightforward
generalization of the one-family expression (\ref{Hsing}).  Given
that there are $N_a$ particles in the $a$th family, the densities
$\rho_a(x)$ must be normalized according to
\beq\label{conservation2F} \int\limits_{-\infty}^\infty\, dx\,
\rho_{1}(x) = N_{1}\,,\quad\quad \int\limits_{-\infty}^\infty\, dx
\rho_{2}(x) = N_{2}\,. \eeq As in the one-family case, these
normalization conditions are implemented by adding to
(\ref{Hcollective2F}) the chemical-potential terms
$\sum_{a=1,2}\mu_a\left(\int\limits_{-\infty}^\infty\,dx \,\rho_a
(x) - N_a\right)$\,.

As was discussed in \cite{BFM}, the collective Hamiltonian
(\ref{Hcollective2F}) is invariant under an Abelian group of
strong-weak-coupling dualities, which is a generalization of the
single-family case (\ref{duality}). A remarkable consequence of
these duality symmetries (see Section 3.1 of \cite{BFM} for more
details) is that when one sets \beq\label{SOHI} \lambda_1\lambda_2
=1\,,\quad\quad \lambda_{12} = -1\eeq in (\ref{threebody}), the
two-family CM (\ref{Hcollective2F}) becomes similar, in some
sense, at the level of collective field theory, to the original
single family CM, with a collective Hamiltonian effectively given by
(\ref{Hcollective}), for a single effective density. More
precisely, this similarity manifests itself in the fact that at
the special point (\ref{SOHI}), the original Hamiltonian
(\ref{Hcollective2F}) can be mapped by these duality symmetries onto
a two-family collective Hamiltonian in which the two families are
still {\em distinct}, but have common mass and two-body
interaction couplings, and therefore, the two densities can be
combined into a certain effective one-family density $\rho_{eff}$.
In fact, at these special points, the classical densities (i.e.,
the static solutions $\rho_1(x)$ and $\rho_2(x)$ of the equations
of motion associated with (\ref{Hcollective2F})) turn out to be
proportional to each other, and of opposite signs. Thus, for
example, for $m_2 = -{m_1\over\lambda_1} < 0$, the common
parameters mentioned above are $\lambda=\lambda_1$ and $m=m_1$,
leading to an effective one-family density \beq\label{rhoeff}
\rho_{eff} = \rho_1 -{\rho_2\over\lambda_1}\,,\eeq which satisfies
the static equation of motion of the single-family model
(\ref{Hcollective}) with these common parameters, whereas for $m_1
= -{m_2\over\lambda_2} < 0$, one obtains similar relations, but
with the two families interchanged. (Negative masses and densities
in these formulas are interpreted as quantities corresponding to
holes rather than particles, as was mentioned above.)

In conclusion of this introduction it is proper to mention that
the Heisenberg equations of motion of the collective field $\rho
(x)$ and its conjugate momentum $\pi (x)$ may be interpreted as
the isentropic hydrodynamic flow equations of an Eulerian fluid
\cite{hydro} (see also  \cite{JFnmm} ) and the latter may be associated with
the completely integrable and soliton-bearing Benjamin-Ono
equation, both at the classical level \cite{Jevicki:1991yi, abanov1} and
the quantum level \cite{abanov}.

This paper is organized as follows: In Section 2 we solve the
static BPS equation associated with the one-family collective
Hamiltonian (\ref{Hcollective}) by converting it into a Riccati
equation which we then solve explicitly. The solution is a static
periodic density wave - the finite amplitude wave solution of
\cite{Polychronakos:1994xg}. Conversion of the BPS equation into a
Riccati equation is achieved by considering the complex valued
resolvent  $\Phi(z)$ associated with the positions of the $N$
particles on the line (see Eq.(\ref{Phi})), whose boundary value,
as the complex variable $z$ approaches the real axis, is a linear
combination of the density field $\rho (x)$ and its
Hilbert-transform $\rho^H(x)$ (see Eq. (\ref{Phipm})) \cite{ajj2,
abanov}. That the latter combination satisfies the Riccati
equation then follows from the BPS-equation and its
Hilbert-transform. We then study various limits of the periodic
solution. We conclude Section 2 by showing that the coupled
BPS-equations, associated with (\ref{Hcollective2F}) at the
special point (\ref{SOHI}) in parameter space do indeed collapse
into a single-family BPS equation.

In Section 3 we consider the static limit of the equation of
motion associated with (\ref{Hcollective}) - namely, the full
variational equation. Every solution of the BPS-equation is, of
course, a solution of the full variational equation. It is more
challenging to find non-BPS solutions of the latter. We seek such
solutions in the form of BPS configurations shifted by a constant,
as was mentioned above. For each of the cases $\lambda >1 $ and $0
< \lambda <1$ we find two types of solutions, namely, a positive
periodic density wave (a vortex crystal) and a negative one (an anti-vortex crystal).
We discuss how these solutions map onto each other by the duality
transformations (\ref{duality}). Then, we discuss the energy
density of these non-BPS solutions, averaged over a period. We end
Section 3 by showing that the coupled variational equations,
associated with (\ref{Hcollective2F}) at the special point
(\ref{SOHI}) in parameter space  collapse into a
single-family variational equation.

For the sake of completeness, and also for future use, we provide and
prove in Appendix A a compendium of useful identities involving
Hilbert-transforms.

In Appendix B we note and also resolve a mathematical paradox
associated with the variational equation. It has to do with the
trilocal term in the density fields obtained by expanding the square
in (\ref{Hcollective}). In many papers on the collective approach
to the Calogero model, that trilocal term is converted into a
local $\rho^3 (x)$ term by employing a certain identity among
distributions, Eq.(\ref{PPdelta}). However, strictly speaking,
that identity is valid only for distributions acting on test
functions which are integrable along the whole real line. The
periodic density profiles discussed in this paper are certainly
not of this type. Nevertheless, they arise correctly as solutions
of the variational equation associated with the alternative form
of the collective potential containing the $\rho^3 (x)$ term,
given in Eq.(\ref{Vcoll1}), as they do, for example, in the
pioneering work \cite{Polychronakos:1994xg}, where these periodic
density waves were discovered. The resolution of this paradox lies in
proper readjustment of the chemical potential enforcing the
constraint (\ref{conservation}).

Finally, in order for this paper to be self-contained, we briefly 
summarize in Appendix C the derivation of the collective-field Hamiltonian (\ref{Hcollective}) from (\ref{Hcalogero}).


\section{Periodic BPS Density Waves: Soliton Crystals}
The Hamiltonian (\ref{Hcollective}) is essentially the sum of two
positive terms\footnote{\label{fn5}Recall the constraint (\ref{conservation}) and our 
comment concerning $H_{sing}$ following  (\ref{conservation}). In addition, as was mentioned above, the external
confining potential was set to zero. Thus, the first two terms in
(\ref{Hcollective}) comprise the BPS limit of the model.}. Its
zero-energy classical solutions are zero-momentum, and therefore
time independent configurations of the collective field
(\ref{collective}), which are also solutions of the BPS equation
\beq\label{BPS} B[\rho] \equiv \frac{\lambda - 1}{2}
\frac{\partial_{x} \rho}{\rho} + \lambda \pv \int \frac{ dy
\rho(y)}{x - y} = 0\,. \eeq

It is easy to check that the duality transformation
(\ref{duality}) maps a solution $\rho(x)$ of (\ref{BPS}) with
coupling $\lambda$ onto another solution $\tilde\rho(x) =
-\lambda\rho(x)$ of that equation with coupling $\tilde\lambda =
{1\over\lambda}$. As we shall see below in Eq. (\ref{rhorhoH}),
all solutions of (\ref{BPS}) are of definite sign, and never
vanish along the real axis. Thus, such a positive solution of
(\ref{BPS}) is mapped by (\ref{duality}) onto a negative solution,
and vice-versa.

The BPS equation (\ref{BPS}) may be written alternatively as
\beq\label{BPS1} (\lambda -1)\,\partial_x\rho =
2\pi\lambda\rho\rho^H\,,\eeq where $\rho^H$ is the
Hilbert-transform (\ref{Hilbert}) of $\rho$. Note that for
$\lambda = 1$, where the CM describes non-interacting fermions\footnote{
The constant solution is also the sole solution of  (\ref{BPS1}) when $\lambda =0$, corresponding to 
non-interacting bosons.}
the only solution of (\ref{BPS1}) is $\rho = \rho_0 = {\rm
const.}$ Henceforth, we shall assume $\lambda\neq 1$.

The proper way to solve this nonlinear integro-differential
equation is to consider it together with its
Hilbert-transform\cite{ajj2, abanov} \beq\label{BPS-H} (\lambda
-1)\,\partial_x\rho^H = \pi\lambda ((\rho^H)^2 - \rho^2 +
\rho_0^2)\,,\eeq where on the RHS we used the identity
(\ref{ffHilbert}) (and the fact that $\partial_x\rho^H =
(\partial_x\rho)^H$ on the LHS). Here $\rho_0$ is a real parameter
such that \beq\label{subtraction} \int\limits_{-\infty}^\infty\,
dx\, (\rho(x) - \rho_0) = 0\,.\eeq It arises from the fact that we
seek a solution of $\rho (x)$ which need not necessarily decay at
spatial infinity. (See (\ref{rhobar} ).) Note that (\ref{BPS-H}) is even in $\rho_0$. By
definition, the sign of $\rho_0$ coincides with that of $\rho
(x)$, the solution of (\ref{BPS-H}). A positive solution $\rho
(x)\geq 0$ corresponds to a BPS configuration of particles, and a
negative one - to a configuration of antiparticles, as was
mentioned following (\ref{duality}).

To arrive at the Riccati equation mentioned in the introduction,
we proceed as follows. Given the density $\rho(x)$, consider the
resolvent \beq\label{Phi} \Phi(z) =
-{1\over\pi}\int\limits_{-\infty}^\infty \,dy\,{ \rho(y)\over z-y}
\eeq associated with it, in which $z$ is a complex variable. It is
easy to see that it is related to the resolvent $G(z)$ of the {\em
subtracted} density $\bar\rho(x) = \rho (x) - \rho_0$, defined in
(\ref{G}), by \beq\label{PhiG} \Phi (z) = -{1\over\pi}\, G(z) +
i\rho_0\,\sign (\Im z)\,.\eeq

The resolvent $\Phi(z)$ is evidently analytic in the complex
plane, save for a cut along the support of $\rho (x)$ on the real
axis. From the identity (\ref{PS}) we obtain \beq\label{Phipm}
\Phi_\pm (x) \equiv \Phi(x\pm i0) = \rho^H(x) \pm i\rho(x) \,,
\eeq consistent with (\ref{PhiG}) and (\ref{G1}). Thus, if $\Phi
(z)$ is known, $\rho (x)$ can be determined from the discontinuity
of $\Phi (z)$ across the real axis.

An important property of $\Phi(z)$, which follows directly from
the definition (\ref{Phi}), is \beq\label{herglotz0} \Im\, \Phi(z)
= {\Im\,z\over\pi}\,\int\limits_{-\infty}^\infty \,
{\rho(y)\,dy\over |z-y|^2}\,. \eeq Thus, if $\rho(x)$ does not
flip its sign throughout its support, we have \beq\label{herglotz}
\sign\,\left(\Im\, \Phi(z)\right) =
\sign\,\left(\Im\,z \right)\sign\,\left(\rho (x)\right)\,.\eeq We
shall use this property to impose certain further conditions on
the solution of (\ref{Riccati1}) below.

It follows from (\ref{Phipm}) that (\ref{BPS1}) and (\ref{BPS-H})
are, respectively, the imaginary and real parts of the  Riccati
equation \beq\label{Riccati} (\lambda -1)
\partial_x \Phi_\pm (x) = \pi\lambda (\Phi_\pm^2(x) + \rho_0^2)
\eeq obeyed by both complex functions $\Phi_\pm (x)\,.$ Let
$\Phi_\pm (z)$ be the analytic continuations of $\Phi_\pm (x)$
into the $z-$upper and lower half planes, respectively. These
functions are evidently the two solutions of \beq\label{Riccati1}
(\lambda -1)\partial_z \Phi (z) = \pi\lambda (\Phi (z)^2 +
\rho_0^2)\,, \eeq subjected to the boundary conditions $\Phi^*_+
(x+i0) = \Phi_- (x-i0)$ and $\sign\left(\Im \Phi_+ (x + i0)\right)
= \sign\left(\rho(x)\right) = \sign\rho_0,$ from (\ref{Phipm}).
The resolvent (\ref{Phi}) is then obtained by patching together
$\Phi_+(z)$ in the upper half-pane and $\Phi_-(z)$ in the lower
half-plane.

The standard way to solve (\ref{Riccati1}) is to write it as
\beq\label{Riccati2} \left({1\over \Phi(z) - i\rho_0} - {1\over
\Phi(z) + i\rho_0} \right)\,\partial_z \Phi (z) = i k \,, \eeq
where \beq\label{k} k = {2\pi\lambda\rho_0\over\lambda -1}\,, \eeq
is a real parameter.

Straightforward integration of (\ref{Riccati2}) then yields the
solutions \beq\label{PhiSol} \Phi_\pm (z)  = i\rho_0\,{1 + e^{ikz
-u_\pm}\over 1 - e^{ikz -u_\pm}}\,,\eeq where $u_\pm$ are
integration constants. The boundary condition $\Phi_+^* (x+i0) =
\Phi_- (x-i0)$ then tells us that $u_- = -u_+^*$. Clearly, $\Im
u_+$ can be absorbed by a shift in $x$. Therefore, with no loss of
generality we set $\Im u_+ = 0$. The second boundary condition
$\sign\left(\Im \Phi_+ (x+i0)\right) = \sign\rho_0 $ then tells us
that $u \equiv \Re u_+ >0\,.$ Thus, $\Phi_\pm (z)$ are completely
determined and we obtain (\ref{Phi}) as \beq\label{PhiSolFinal}
\Phi (z)  = i\rho_0\,{1 + e^{ikz -u\,\sign(\Im z)}\over 1 - e^{ikz
-u\,\sign(\Im z)}}\,.\eeq As can be seen in (\ref{rhorhoH}) below,
the density $\rho(x)$ associated with (\ref{PhiSolFinal}) is
indeed of definite sign, namely, $\sign\rho_0$.

The asymptotic behavior of (\ref{PhiSolFinal}) is such that
\beq\label{asymptotics} \Phi (\pm i\infty ) = \pm i\rho_0\,\sign
k\,. \eeq This must be consistent with (\ref{herglotz}), which
implies (together with the fact that $\sign\left(\rho(x)\right) =
\sign\rho_0$) that $k$ must be {\em positive}. In other words, as
can be seen from (\ref{k}), positive (space-dependent) 
BPS density configurations
($\rho_0 > 0 $) exist only for $\lambda
>1$, and negative (space-dependent) BPS densities ($\rho_0 < 0 $) arise only for
$0<\lambda<1$\footnote{Constant solutions $\rho = \rho_0$ of (\ref{BPS}), are of course not subjected to this
correlation between $\sign \rho_0$ and the range of $\lambda$.} . The duality symmetry (\ref{duality}), which
interchanges the domains $0< \lambda <1$ and $\lambda >1$, maps
these two types of BPS configurations onto each other.

Now that we have determined $\Phi (z)$, let us extract from it the
BPS density $\rho (x)$ and its Hilbert transform $\rho^H(x)$. From
(\ref{PhiSolFinal}) we find that \beq\label{PhiSolplus} \Phi_+(x)
= \Phi(x+i0) = \rho_0\, {-\sin\, kx + i\sinh u\over \cosh u -\cos
kx}\,,\eeq from which we immediately read-off the solution of the
BPS-equation (\ref{BPS}) as \beqra\label{rhorhoH} \rho (x) &=&
\,\,\,\,\rho_0\, {\sinh u\over \cosh u -\cos
kx}\nonumber\\{}\nonumber\\\rho^H(x) &=&  - \rho_0\, {\sin kx
\over \cosh u -\cos kx}\,,\eeqra where both $k > 0$ and $u>0$, and
the sign of $\rho(x)$ coincides with that of $\rho_0$. That
$\rho^H$ in (\ref{rhorhoH}) is indeed the Hilbert-transform of
$\rho$ can be verified by explicit calculation.

The static BPS density-wave, given by $\rho(x)$ in (\ref{rhorhoH}), is
nothing but the finite-amplitude solution of
\cite{Polychronakos:1994xg}. It comprises a two-parameter family
of spatially periodic solutions, all of which have zero energy
density, by construction. The period is \beq\label{period} T =
{2\pi\over k} = {\lambda -1\over\lambda\rho_0}\,.\eeq It can be
checked by explicit calculation\footnote{The best way to do this
computation is to change variables to $t = e^{ikx}$ and transform
the integral into a contour integral around the unit circle.} that
\beq\label{period-av} {1\over T}\int\limits_{\rm period}\, \rho
(x)\, dx = \rho_0\,,\eeq and therefore that
$\int\limits_{-\infty}^\infty\, (\rho (x) - \rho_0) \, dx = 0\,,$
as required by definition of $\rho_0$. Thus, the parameter
$\rho_0$ determines both the period of the solution $\rho (x)$, as
well as its period-average, and the other (positive) parameter $u$
determines the amplitude of oscillations about the average value.
Note also from (\ref{period-av}), that the number of particles per
period is \beq\label{ppp} T\rho_0 = {\lambda-1\over
\lambda}\,.\eeq

A couple of limiting cases of (\ref{rhorhoH}) are worth
mentioning. Thus, if we let $u\rightarrow 0\,,$ we obtain a comb
of Dirac $\delta-$functions \beq\label{comb} \rho (x) =
{\lambda-1\over \lambda}\, \sum\limits_{n\in Z\!\!\!Z}\, \delta (x
- nT)\,.\eeq If, in addition to $u\rightarrow 0$, we also let $k$
tend to zero (or equivalently, let the period $T$ diverge), such
that $b = {u\over k}$ remains finite, we obtain the BPS soliton
solution \cite{Polychronakos:1994xg, Andric:1994nc}
\beq\label{lump} \rho (x) = {\lambda-1\over
\lambda}\,{1\over\pi}\, {b\over b^2 + x^2}\,.\eeq In fact, the
original construction of the periodic soliton (\ref{rhorhoH}) in
\cite{Polychronakos:1994xg} was done by juxtaposing infinite
solitons like (\ref{lump}) in a periodic array. For this reason we may refer to the finite amplitude
BPS density wave in (\ref{rhorhoH}) also as the {\em soliton crystal}.

Note that the relation (\ref{ppp}) is preserved in both limiting
cases, since the RHS of (\ref{ppp}) depends neither on $u$ nor on
$k$.

\subsection{BPS Solutions of the Two-Family Model at the
Special Point (\ref{SOHI})} The BPS-equations of the two-family
collective field Hamiltonian (\ref{Hcollective2F}) are
\beqra\label{BPS2F} B_1[\rho_1,\rho_2] &\equiv& \frac{\lambda_{1}
- 1}{2} \frac{\partial_{x} \rho_{1}}{\rho_{1}} + \lambda_{1} \pv
\int \frac{ dy\rho_{1}(y)}{x - y}  + \lambda_{12}
\pv \int \frac{dy\rho_{2}(y)}{x - y}  = 0\nonumber\\{}\nonumber\\
B_2[\rho_1,\rho_2] &\equiv& \frac{\lambda_{2} - 1}{2}
\frac{\partial_{x} \rho_{2}}{\rho_{2}} + \lambda_{2} \pv \int
\frac{ dy\rho_{2}(y)}{x - y} + \lambda_{12} \pv \int
\frac{dy\rho_{1}(y)}{x - y} = 0\,.\eeqra Solutions of these
coupled equations yield the time-independent zero-energy and
zero-momentum configurations of the collective fields $\rho_1$ and
$\rho_2$.

Finding the general solution of these coupled equations for
arbitrary couplings and masses (subjected to (\ref{threebody})) is
still an open problem, which we do not address in the present
paper. However, at the special point (\ref{SOHI}), where the
two-family model becomes similar to a single-family model, the two
equations (\ref{BPS2F}) simplify drastically, becoming linearly
dependent. For example, for \beq\label{SOHI1} \lambda = \lambda_1
= {1\over\lambda_2}\,,\quad \lambda_{12} = -1\,,\quad\quad {\rm
and}\quad\quad m=m_1 = -\lambda m_2\,, \eeq it is easy to see that
\beq\label{lindep} B_1 + \lambda B_2 = {\lambda -1\over 2}\,
\partial_x\,\log\,\left({\rho_1\over\rho_2}\right)\,. \eeq
Since at the same time, from (\ref{BPS2F}), $B_1=B_2 = 0$,
(\ref{lindep}) implies that the two densities must be proportional
\beq\label{proportionality} \rho_2(x) = -\kappa \rho_1(x)\,.\eeq
(From the discussion in \cite{BFM} we know that the constant
$\kappa >0$, and the negative density is interpreted as density of
holes, as was mentioned in the Introduction.) Upon substitution of
(\ref{proportionality}) back in (\ref{BPS2F}) we see that
\beq\label{B1} B_1 = \frac{\lambda - 1}{2} \frac{\partial_{x}
\rho_{eff}}{\rho_{eff}} -\lambda\pi \rho_{eff}^H\,, \eeq coincides
with the corresponding one-family expression $B$ in (\ref{BPS})
with an effective density $\rho_{eff}$ given by (\ref{rhoeff}).
Thus, at this special
point, $\rho_{eff} (x)$ is given by (\ref{rhorhoH}), from which
$\rho_1$ and $\rho_2$, being proportional to $\rho_{eff}$, can be
deduced as well. An analogous solution of (\ref{BPS2F}) exists for
the case in which the roles of the two families in (\ref{SOHI1})
are interchanged.


\section{Non-BPS Solutions of the Equation of Motion}
The uniform-density ground state, as well as the periodic
space-dependent BPS-configurations discussed in the previous
section, all correspond to zero-energy and zero-momentum
configurations of the collective field Hamiltonian
(\ref{Hcollective}). Static density configurations with {\em positive}
energy density are found by extremizing the collective potential
\beqra\label{Vcoll} V_{coll} &=& \frac{1}{2 m} \int dx\, \rho(x)
{\left( \frac{\lambda - 1}{2} \frac{\partial_{x} \rho}{\rho} +
\lambda \pv \int \frac{ dy \rho(y)}{x - y} \right)}^{2} +
\mu\left(N - \int\,dx \,\rho (x) \right)\nonumber\\{}\nonumber\\
&=&  \frac{1}{2 m} \int dx\, \rho(x) B[\rho]^2 + \mu\left(N -
\int\,dx \,\rho (x) \right)\eeqra part of (\ref{Hcollective}).

Computation of the variation of (\ref{Vcoll}) with respect to
$\rho$ is most easily carried with the help of (\ref{Bvar}) just
below. Thus, using the elementary relation $\int\,dx\,F(x)\,G^H(x)
= - \int\,dx\,F^H(x)\,G(x)$ it is easy to obtain the variational
identity \beq\label{Bvar} \int\,dx\,\rho(x)\,F(x)\,\delta B[\rho]
= \int\,dx\,\left[-{\lambda -1\over 2\rho}\,\partial_x (\rho F) +
\pi\lambda\,(\rho F)^H \right]\delta\rho(x)\,,\eeq where the
infinitesimal variation of $B[\rho]$  \beq\label{deltaB} \delta
B[\rho] = {\lambda -1\over 2}\,\partial_x
\left({\delta\rho\over\rho}\right) - \pi\lambda\,\delta\rho^H \eeq
was computed from (\ref{BPS}).

Using (\ref{Bvar}), it is straightforward to obtain the desired
variational equation as \beq\label{variationaleq} 2m\,{\delta
V_{coll}\over\delta \rho(x)} = B[\rho]^2 - {\lambda
-1\over\rho}\,\partial_x (\rho B[\rho]) + 2\pi\lambda\,(\rho
B[\rho])^H - 2m\mu = 0\,.\eeq

The collective potential (\ref{Vcoll}) is invariant under the
duality transformation (\ref{duality}). Thus,
(\ref{variationaleq}) must transform covariantly under
(\ref{duality}). Indeed, it is straightforward to see that under
(\ref{duality}), the variational equation (\ref{variationaleq})
transforms into ${1\over \lambda^2}$ times itself. In this way, a
solution $\rho(x)$ of (\ref{variationaleq}) with parameters
$\lambda, m, \mu$ is transformed into a solution $\tilde\rho (x)$
of (\ref{variationaleq}) with parameters $\tilde\lambda, \tilde m,
\tilde\mu$. We shall make use of this fact later-on.

Evidently, any solution of the BPS equation $B[\rho] = 0$
(Eq.(\ref{BPS})) is also a solution of the variational equation
(\ref{variationaleq}) (with $\mu=0$), reflecting the fact that
(\ref{Vcoll}) is quadratic and homogeneous in $B[\rho]$.

Unfortunately, we do not know how to find the most general
solution of this equation. Therefore, we shall content ourselves
with finding a particular family of solutions to
(\ref{variationaleq}) of the simple shifted form
\beq\label{ansatz} \rho (x) = \rho_s(x) + c\,, \eeq where
\beq\label{rhos} \rho_s (x) = \rho_0\, {\sinh u\over \cosh u -\cos
kx} \eeq is the BPS profile in (\ref{rhorhoH}) and $c$ is an
unknown constant, to be determined from (\ref{variationaleq}).
(Clearly, (\ref{ansatz}) with $c=0$ must be a solution of
(\ref{variationaleq}).)

Let us proceed in a few steps. First, note that \beq\label{rhoH}
\rho^H(x) = \rho_s^H(x) = - \rho_0\, {\sin kx \over \cosh u -\cos
kx}\,, \eeq from (\ref{rhorhoH}). Then, compute \beq\label{Brhos}
B[\rho] = \frac{\lambda - 1}{2} \frac{\partial_{x} \rho}{\rho} -
\lambda\pi\rho^H = {\lambda\pi c\rho_0\sin\,kx\over c(\cosh\,u -
\cos\,kx) + \rho_0\sinh u}\,, \eeq from which we obtain the
remarkably simple relation \beq\label{rhoBrho} \rho\,B[\rho] =
-\lambda\pi c \rho^H\,.\eeq Therefore, \beq\label{rhoBrhoH}
(\rho\,B[\rho])^H  = \lambda\pi c (\rho_s - \rho_0)\,,\eeq where
we used the identity (\ref{HilbertSq}). (Note that (\ref{rhoBrhoH}) is consistent with the
identity $\int_{-\infty}^\infty\,F^H(x)\,dx = 0$.) Substituting the ansatz
(\ref{ansatz}) and the auxiliary results
(\ref{Brhos})-(\ref{rhoBrhoH}) in (\ref{variationaleq}) we obtain
the LHS of the latter as a rational function of polynomials in
$\cos\,kx$. The numerator of that function is a cubic polynomial,
which we then expand into a finite cosine Fourier series, all
coefficients of which must vanish. Thus, the coefficient of $\cos
3kx$ determines the chemical potential in terms of the remaining
parameters as \beq\label{mu} \mu = -{(\lambda\pi)^2\over
2m}\,\rho_0(\rho_0 + 2c)\,,\eeq which we then feed into the
coefficients of the remaining three terms. The coefficient of
$\cos 2kx$ is then found as the cubic \beq\label{cubic}
-(\lambda\pi)^2\rho_0\,c (c^2\sinh\,u + 2c\rho_0\cosh\,u +
\rho_0^2\sinh\,u)\,,\eeq where we used (\ref{k}) on the way. This
coefficient must vanish, yielding a cubic equation for $c$. The
remaining Fourier coefficients vanish identically upon
substitution of the roots of this cubic equation for $c$.

As we have anticipated following (\ref{ansatz}), one root of this
cubic equation is obviously $c_0=0$, which corresponds to $\rho =
\rho_s$. The other two roots are \beq\label{c12} c_1 =
-\rho_0\,\tanh\,{u\over 2}\quad\quad {\rm and}\quad\quad c_2  =
-\rho_0\,\coth\,{u\over 2}\,. \eeq Note that neither of these
roots, and therefore neither of the shifted solutions
(\ref{ansatz}), depend on $m$ or on $\mu$. Once the parameters $m$
and $\mu$ are related according to (\ref{mu}), they drop out of any further
consideration.

\subsection{Large Amplitude Density Waves: Vortex Crystal Solutions}
From this point onward we shall discuss the cases $\lambda > 1$
and $0 < \lambda <1$ separately.

\subsubsection{The case $\lambda >1$} In this case $\rho_0 >0$, as
we saw following (\ref{asymptotics}). For positive $\rho_0$, the
first root $c_1$ in (\ref{c12}) amounts in (\ref{ansatz}) to
shifting the BPS solution $\rho_s(x)$ by its {\em minimum}. The
resulting solution \beq\label{LAW} \rho_p (x) = \rho_0\,
\left({\sinh u\over \cosh u -\cos kx} - \tanh\,{u\over
2}\right)\eeq is a positive function which vanishes periodically.
We shall refer to it as the {\em vortex crysta}l solution, as it is a
periodic generalization of the single vortex solution of
\cite{Andric:1994nc}. Since $\rho_p (x) >0\,,$ it is a density of
particles (rather than holes). Therefore it corresponds to having
a positive mass parameter $m>0$ in (\ref{Vcoll}). The vortex 
crystal (\ref{LAW}) coincides with the so-called {\em large
amplitude} wave solution of \cite{Sen:1997qt} for the case
$\lambda > 1$ and zero velocity.

The second root $c_2$ in (\ref{c12}) amounts to shifting the BPS
solution $\rho_s(x)$ by its {\em maximum}. The resulting solution
\beq\label{negLAW} \rho_n (x) = \rho_0\, \left({\sinh u\over \cosh
u -\cos kx} - \coth\,{u\over 2}\right)\eeq is thus a negative
function which vanishes periodically - an anti-vortex crystal.
Being a negative solution of
(\ref{variationaleq}), Eq.(\ref{negLAW}) should be interpreted as
the density of holes rather than particles. Therefore it
corresponds to having a negative mass $m<0$ in (\ref{Vcoll}).

\subsubsection{The case $0 < \lambda < 1$} In this case $\rho_0 < 0$, as
we saw following (\ref{asymptotics}). Therefore $c_1$ and $c_2$ in
(\ref{c12}) switch roles: For negative $\rho_0$, $c_1$ amounts to
shifting in (\ref{ansatz}) the BPS solution $\rho_s(x)$ by its
{\em maximum}. The resulting solution \beq\label{tnLAW}
\tilde\rho_n (x) = \rho_0\, \left({\sinh u\over \cosh u -\cos kx}
- \tanh\,{u\over 2}\right)\eeq is a negative function which
vanishes periodically - an anti-vortex crystal.
It is therefore a density of holes corresponding to having a
negative mass $m<0$ in (\ref{Vcoll}).

The second root $c_2$ amounts in this case to shifting the BPS
solution $\rho_s(x)$ by its {\em minimum}. The resulting solution
\beq\label{tLAW} \tilde\rho_p (x) = \rho_0\, \left({\sinh u\over
\cosh u -\cos kx} - \coth\,{u\over 2}\right) = |\rho_0|\,
\left(\coth\,{u\over 2} - {\sinh u\over \cosh u -\cos
kx}\right)\eeq is thus a positive function which vanishes
periodically - a vortex crystal. It corresponds to having $m>0$
in (\ref{Vcoll}), in a similar manner to $\rho_p(x)$ in
(\ref{LAW}). $\tilde\rho_p (x)$ coincides with the large amplitude
wave solution of \cite{Sen:1997qt} for the case $0<\lambda < 1$
and zero velocity. Note that $\tilde\rho_p (x)$ has appeared also
in \cite{ajj2}.

The duality transformations (\ref{duality}) leave the wave-number
$k$ in (\ref{k}) invariant. By definition, the positive parameter
$u$, defined in (\ref{PhiSolFinal}), is invariant under
(\ref{duality}) as well. Thus, evidently, the duality
transformations (\ref{duality}) map $\rho_p (x)$ in (\ref{LAW})
and $\tilde\rho_n(x)$ in (\ref{tnLAW}) onto each other. (Of
course, the $\rho_0$ parameters appearing in the latter two
equations are different from each other, and related by the fourth
relation in (\ref{duality}).) Similarly, the duality
transformations (\ref{duality}) map $\rho_n (x)$ in (\ref{negLAW})
and $\tilde\rho_p(x)$ in (\ref{tLAW}) onto each other.

\subsubsection{Average Energy Densities per Period}
Our new solutions (\ref{LAW}) - (\ref{tLAW}) of the
variational equation (\ref{variationaleq}) are periodic functions,
with the same period as that of the BPS solution $\rho_s$. Since
these are non-BPS configuration, they must carry positive energy
density\footnote{Since negative densities correspond to
holes, whose mass should be taken negative, we have ${\rho(x)\over m} > 0$ in these
cases as well. This renders
$V_{coll}$ in (\ref{Vcoll}) positive for such densities. Thus, the
negative solutions $\rho_n$ and $\tilde\rho_n$ carry positive
energy density, as their positive counterparts obviously do.}. We
shall now proceed to calculate the mean energy densities per
period of these configurations, to which end we must determine the
combination $\rho B[\rho]^2$ appearing in (\ref{Vcoll}). From the
general expressions (\ref{Brhos}) and (\ref{rhoBrho}) we obtain
\beq\label{rhoBrhosq} \rho B[\rho]^2  = {\lambda\pi
c\rho_0\sin\,kx\over \cosh\,u - \cos\,kx} \,{\lambda\pi
c\rho_0\sin\,kx\over c(\cosh\,u - \cos\,kx) + \rho_0\sinh
u}\,.\eeq The desired period-averaged energy density is then given
by \beq\label{av-energy} {\cal E} = {1\over T}\,\int\limits_{\rm
period}\, dx\, {\rho B[\rho]^2\over 2m}\,,\eeq where $T={2\pi\over
k}$ (Eq. (\ref{period})).

We shall content ourselves with computing the energy density only
of the positive densities $\rho_p$ and $\tilde\rho_p$. In order to
compute ${\cal E}$ of (\ref{LAW}), corresponding to $\lambda >1$
and $\rho_0 >0$, we substitute $c = c_1$ in (\ref{rhoBrhosq}).
After some algebra, we find that in this case
\beq\label{rhoBrhosqp} \rho_p B[\rho_p]^2 =
(\lambda\pi\rho_0)^2\,\tanh{u\over 2}\,\left[\rho_0
(1-\tanh{u\over 2}) - (\rho_s(x) - \rho_0) \tanh{u\over
2}\right]\,.\eeq In view of (\ref{period-av}), and by definition
of $\rho_0$, the period-average of $\rho_s(x) - \rho_0$ vanishes.
Thus, from (\ref{Vcoll}), we obtain the period-average energy
density of (\ref{LAW}) as \beq\label{energyp} {\cal E}[\rho_p] =
{(\lambda\pi\rho_0)^2\over 2m}\,\rho_0\,\tanh{u\over
2}(1-\tanh{u\over 2})\,,\eeq which is a manifestly positive
quantity. It depends continuously on the two parameters $\rho_0$
and $u$, comprising an unbounded continuum of positive energies,
which is not gapped from the zero energy density of the BPS
solitons.

Similarly, in order to compute ${\cal E}$ of (\ref{tLAW}),
corresponding to $0 < \lambda < 1$ and $\rho_0 < 0$, we substitute
$c = c_2$ in (\ref{rhoBrhosq}). After some algebra, we find that
in this case \beq\label{rhoBrhosqn} \tilde\rho_p B[\tilde\rho_p]^2
= -(\lambda\pi\rho_0)^2\,\coth{u\over 2}\,\left[\rho_0
(\coth{u\over 2} -1 ) + (\rho_s(x) - \rho_0) \coth{u\over
2}\right]\,,\eeq which leads to the positive period-average energy
density of (\ref{tLAW}) given by \beq\label{energyn} {\cal
E}[\tilde\rho_p] = -~ {(\lambda\pi\rho_0)^2 \over
2m}\,\rho_0\,\coth{u\over 2}(\coth{u\over 2}-1)\,.\eeq

\subsubsection{Energy Densities at Fixed Average Particle Density}
It is particularly useful to consider the energy densities
(\ref{energyp}) and (\ref{energyn}) at a fixed average particle
density per period. The latter is, of course, the subtraction
constant as defined in (\ref{subtraction}), which is given by
\beq\label{trho0}\tilde\rho_0 = \rho_0 + c \eeq for the shifted
solutions (\ref{ansatz}). Both (\ref{energyp}) and (\ref{energyn})
depend on the two parameters $\rho_0$ and $u$. Holding
$\tilde\rho_0$ fixed can thus be used to eliminate one of these
parameters, which we shall take to be $u$.

Let us concentrate first on $\rho_p$ in (\ref{LAW}), for which $c=
c_1 = -\rho_0\,\tanh\,{u\over 2}$ (and of course, $\lambda >1$).
Thus, \beq\label{postilderho0} \tilde\rho_0 = \rho_0\,(1 -
\tanh\,{u\over 2})\,,\eeq which is positive, since $\rho_0
> 0$ in (\ref{LAW}). Moreover, $\rho_0\geq\tilde\rho_0$ in this case,
since $u>0$. In terms of this fixed $\tilde\rho_0$, we obtain
${\cal E}[\rho_p]$ in (\ref{energyp}) as \beq\label{energyp1}
{\cal E}[\rho_p] = {(\lambda\pi)^2\over
2m}\,\tilde\rho_0\rho_0\,(\rho_0 - \tilde\rho_0)\,,\quad
\rho_0\geq \tilde\rho_0 = {\rm fixed}\,.\eeq This energy density
vanishes at the minimal possible value of $\rho_0 = \tilde\rho_0$,
corresponding to $u=0$, and therefore to the BPS density
configuration (\ref{comb}). As $\rho_0$ increases from its minimal
value, the period-average energy density ${\cal E}[\rho_p]$
increases monotonically from zero to infinity. Increasing $\rho_0$
really means increasing the wave number $k =
{2\pi\lambda\rho_0\over\lambda -1}$, i.e., making the density
modulation wave-length shorter.

It is interesting to note that in terms of $k$ and $\tilde\rho_0$
we can write \beq\label{energyp2} {\cal E}[\rho_p] =
{(1-\lambda)\tilde\rho_0\over 4}\,\left({1-\lambda \over 2m}\,k^2
+ {\lambda\pi\tilde\rho_0\over m}k\right)\,,\quad k =
{2\pi\lambda\over \lambda -1}\rho_0 \geq {2\pi\lambda\over \lambda
-1}\tilde\rho_0 = {\rm fixed}\,,\eeq where the expression within
the brackets is nothing but the dispersion relation for
fluctuations around the constant background $\tilde\rho_0$
\cite{Andric:1994su}.

We can analyze the periodic vortices $\tilde\rho_p$ in
(\ref{tLAW}) in a similar manner. For these solutions $c= c_2 =
-\rho_0\,\coth\,{u\over 2}$ (and of course $0<\lambda<1$). Thus,
\beq\label{negtrho0} \tilde\rho_0 = \rho_0\,(1 - \coth\,{u\over
2}) = |\rho_0|\,(\coth\,{u\over 2} - 1)\,,\eeq which is again
positive, since the allowed range of $\rho_0$ in (\ref{tLAW}) is
$\rho_0 \leq 0$. For a given value of $\tilde\rho_0$, $\rho_0 =
-\frac{1}{2} (e^u-1)\tilde\rho_0$ ranges throughout the negative real
axis as $u$ ranges throughout the positive one. In terms of this
fixed $\tilde\rho_0$, we obtain an expression for ${\cal
E}[\tilde\rho_p]$ in (\ref{energyn}) which coincides with the RHS
of (\ref{energyp1}), but where now $\rho_0 \leq 0$, of course:
${\cal E}[\tilde\rho_p] = {(\lambda\pi)^2\over
2m}\,\tilde\rho_0|\rho_0|\,(|\rho_0| + \tilde\rho_0)\,.$ This
energy density vanishes at the maximal possible value of $\rho_0 =
0$, corresponding to $u=0$, and therefore to the BPS density
configuration (\ref{lump}). As $\rho_0$ becomes increasingly
negative the period-average energy density ${\cal
E}[\tilde\rho_p]$ increases monotonically from zero to infinity.
In terms of the wave number $k$ and $\tilde\rho_0$, we obtain that
${\cal E}[\tilde\rho_p] = {(1-\lambda)\tilde\rho_0\over
4}\,\left({1-\lambda \over 2m}\,k^2 + {\lambda\pi\tilde\rho_0\over
m}k\right)\,,$ which coincides with (\ref{energyp2}), but where
now $k\geq 0$ for any value of $\tilde\rho_0>0$.

\subsection{The Two-Family model at the
Special Point (\ref{SOHI})} The variational equations associated
with the two-family collective potential part of the two-family
collective Hamiltonian (\ref{Hcollective2F}) are
\beqra\label{variational2F} B_1^2 -
{\lambda_1-1\over\rho_1}\partial_x\,(\rho_1 B_1) +
2\lambda_1\pi\,(\rho_1 B_1)^H + 2{m_1\over
m_2}\,\lambda_{12}\pi\,(\rho_2 B_2)^H - 2m_1\mu_1 &=&
0\nonumber\\{}\nonumber\\ B_2^2 -
{\lambda_2-1\over\rho_2}\partial_x\,(\rho_2 B_2) +
2\lambda_2\pi\,(\rho_2 B_2)^H + 2{m_2\over
m_1}\,\lambda_{12}\pi\,(\rho_1 B_1)^H - 2m_2\mu_2 &=&
0\,,\nonumber\\{} \eeqra in straightforward analogy with
(\ref{variationaleq}), where the BPS combinations $B_1$ and $B_2$
were defined in (\ref{BPS2F}).

As in the case of the BPS equations (\ref{BPS2F}), the general
solution of these coupled equations for arbitrary couplings and
masses (subjected to (\ref{threebody})) is still an open problem,
which we do not address in the present paper. However, at the
special point (\ref{SOHI}), where the two-family model becomes
similar to a single-family model, the two equations
(\ref{variational2F}) simplify drastically, becoming linearly
dependent, in much the same way that the BPS equations
(\ref{BPS2F}) got simplified.

Consider, for example, the case (\ref{SOHI1}). In this case
(\ref{lindep}) still holds, of course, but now neither $B_1$ nor
$B_2$ need vanish. Thus, we cannot conclude that $\rho_1$ and
$\rho_2$ must be proportional. Instead, we shall now show that
under the condition (\ref{SOHI1}), there is a non-BPS solution of
the coupled equations (\ref{variational2F}) in which the two
densities are proportional to each other, as in
(\ref{proportionality}). In this case it follows from
(\ref{lindep}) that $B_1 + \lambda B_2 = 0\,.$ Substituting this
relation as well as (\ref{SOHI1}) in (\ref{variational2F}) , we
see that the two equations coincide, provided $\mu_1 +
\lambda\mu_2 =0$, and that their common form is nothing but the
variational equation (\ref{variationaleq}) of the single-family,
for an effective density (\ref{rhoeff}). Thus, at this special
point, $\rho_{eff} (x)$ is given by (\ref{rhos}), (\ref{LAW}) or
(\ref{negLAW}), from which $\rho_1$ and $\rho_2$, being
proportional to $\rho_{eff}$ can be deduced as well. An analogous
solution of (\ref{variational2F}) exists for the case in which the roles
of the two families in (\ref{SOHI1}) are interchanged.

\bigskip
\bigskip

\pagebreak

%


\newpage
\setcounter{equation}{0}
\setcounter{section}{0}
\renewcommand{\theequation}{A.\arabic{equation}}
\renewcommand{\thesection}{Appendix A:}
\section{A Compendium of Useful Hilbert-Transform Identities}
\vskip 5mm
\setcounter{section}{0}
\renewcommand{\thesection}{A}
For the sake of completeness, and also for future reference, in
this Appendix we list and prove some well-known and useful
identities involving Hilbert transforms. 

Consider the class of (possibly complex) functions $\rho (x)$ on
the whole real line $-\infty < x < \infty$, whose Hilbert
transforms \beq\label{Hilbert} \rho^H(x) = {1\over \pi}
\pv\int_{-\infty}^\infty \,dy\, {\rho (y) \over y - x} \eeq exist,
and which can be made integrable by subtracting a constant
$\rho_0$. Let us denote \beq\label{rhobar} \bar\rho (x) = \rho (x)
- \rho_0\,. \eeq (If $\rho (x)$ is already integrable, then
$\rho_0 = 0$, of course.) Thus, for example, if $\rho (x)$ is
periodic with period $T$, with a Fourier zero-mode $\rho_0$, then
$\int\limits_{-\infty}^\infty dx\,\bar\rho (x) =
\int\limits_{-\infty}^\infty dx\,(\rho (x) - \rho_0) = 0$\,.

Given $\bar\rho(x)$, consider the resolvent \beq\label{G} G(z) =
\int\limits_{-\infty}^\infty \,dy\,{ \bar\rho(y)\over z-y} \eeq
associated with it, in which $z$ is a complex variable. The
resolvent $G(z)$ is evidently analytic in the complex plane, save
for a cut along the support of $\bar\rho (x)$ on the real axis.
From the identity \beq\label{PS} {1\over x\mp i0} = {P\over x} \pm
i\pi\delta (x)\,, \eeq where $P$ denotes the Cauchy principal
value, we then obtain the well-known formula \beq\label{G1} G(x\mp
i0) = -\pi \rho^H(x) \pm i\pi\bar\rho(x) \,, \eeq where in the
term one before last we used the fact that $\bar\rho^H(x) =
\rho^H(x)$. Thus, if $G(z)$ is known, $\bar\rho (x)$ can be
determined from the discontinuity of $G(z)$ across the real axis.

As a nontrivial example consider $\rho (x) = \bar\rho (x) =
e^{ix}$. For this function $\rho^H(x) = i e^{ix} = i\rho (x)$ and
$G(z) = -2\pi i \,\theta (\Im z) \,e^{iz}$. Consequently $G(x-i0)
= 0$ and $G(x+i0) = -2\pi i \, e^{ix}$, in accordance with
(\ref{G1}). As yet another example consider the Cauchy probability
distribution $\rho (x) = \bar\rho (x) = {\gamma\over\pi} {1\over
x^2 + \gamma^2}$. For this function $\rho^H(x) = -{1\over\pi}
{x\over x^2 + \gamma^2}$ and $G(z) = {1\over z + i\gamma \sign
(\Im z)}$. Consequently $G(x\mp i0) = {x\pm i\gamma\over x^2 +
\gamma^2}$, in accordance with (\ref{G1}).

For all functions in this class, as $z\rightarrow\infty$, $G(z)$
tends asymptotically to zero not slower than ${1\over z}$, that is
\beq\label{Gasympt} G(z) {_{\stackrel{\displaystyle\sim}
{z\rightarrow\infty}}\,\, } {\cal O}\left({1\over z}\right)\,.\eeq
If, in addition, all moments $M_n = \int\limits_{-\infty}^\infty
\, dx\, x^n \,\bar\rho (x)\,,\quad (n\geq 0)$ of $\bar\rho (x)$
exist, then $G(z)$ is the moment generating function of $\bar\rho
(x)$, namely, it has the large-$z$ expansion \beqast G(z) =
\sum_{n=0}^\infty {M_n\over z^{n+1}}\,. \eeqast

The analyticity properties of $G(z)$ and the bounds on its
asymptotic behavior at infinity are at the heart of our derivation
of the Hilbert-transform identities to follow.

{\em From this point on, we shall take all functions $\rho (x)$ to
be real.} For real $\rho (x)$, we deduce from (\ref{G1}) that
\beq\label{G2}   \rho^H(x) = -{1\over \pi} \Re G(x\mp i0) \quad
{\rm and}\quad \bar\rho (x) = \pm {1\over\pi} \Im G(x\mp i0)\,. \eeq

As a warm-up exercise, let us prove the well-known fact that
\beq\label{HilbertSq} (\rho^H (x))^H = -\bar\rho (x) = \rho_0 -
\rho (x) \,. \eeq Thus, consider \beqast (\rho^H (x))^H  &=&
(\bar\rho^H (x))^H  = {1\over\pi}\int\limits_{-\infty}^\infty
{P\over y-x}\, \bar\rho^H(y)\, dy \\{}\\ &=&
{1\over\pi}\int\limits_{-\infty}^\infty \,dy \,
\Re\left[\left({P\over y-x} -i\pi\delta (y-x)\right)
\left(\bar\rho^H(y) + i\bar\rho (y)\right)\right]\, dy - \bar\rho
(x) \\{}\\ &=& -{1\over \pi^2} \Re\, \int\limits_{-\infty}^\infty
{G(y+i0)\over y-x +i0}\, dy - \bar\rho (x) \,,\eeqast where in the
last step we used (\ref{PS}) and (\ref{G2}). Let us now prove that
the last integral vanishes, from which (\ref{HilbertSq}) follows.
To this end, complete the contour of integration in the last
integral (namely, the line running parallel to the real axis just
above it) by the infinite semi-circle in the upper half-plane $\Im
z>0$, traversed in the positive sense. Let us denote the closed
contour thus formed by $\gamma$. Due to the asymptotic behavior
(\ref{Gasympt}) of $G(z)$ we can establish the first equality in
$$ \int\limits_{-\infty}^\infty {G(y+i0)\over y-x +i0}\, dy =
\oint_\gamma\,dz\, {G(z)\over z-x}= 0\,,$$ whereas the second
equality follows since the contour $\gamma$ encompasses no singularity.

We shall now prove the important identity \beq\label{fgHilbert}
(\rho_1\rho_2^H + \rho_1^H\rho_2)^H = \rho_1^H\rho_2^H -
\rho_1\rho_2 + \rho_{10}\rho_{20} \eeq obeyed by any two functions
$\rho_1(x)$ and $\rho_2(x)$ in the class of functions considered.
Our first step in proving (\ref{fgHilbert}) is to observe that it
may be written equivalently as \beq\label{fgHilbert1}
(\bar\rho_1\bar\rho_2^H + \bar\rho_1^H\bar\rho_2)^H =
\bar\rho_1^H\bar\rho_2^H - \bar\rho_1\bar\rho_2\,. \eeq Consider
now the contour integral \beq\label{fg-contour} I = \oint_{\cal
C_\infty} {G_1(z) G_2(z)\over z-x}\,{dz\over 2\pi i}\,,\eeq where
$G_k(z)$ is the resolvent corresponding to $\bar\rho_k(x)\,\,
(k=1,2)\,,$ $x\in I\!\! R\,,$ and where ${\cal C_\infty}$ is the
circle of infinite radius, centered at the origin. Due to the
asymptotic behavior (\ref{Gasympt}) of the two resolvents,
evidently \beq\label{fg-contour-null} I = 0\,.\eeq Since
$G_{1,2}(z)$ are analytic off the real axis, we can deform ${\cal
C_\infty}$ into the positively oriented boundary $\Gamma$ of an
infinitesimal strip around the real axis (namely, the union of a
line parallel to the real axis just below it, traversed from
$-\infty$ to $\infty$, with a line parallel to the real axis just
above it and traversed in the opposite direction). The contour
integral around $\Gamma$ essentially picks up the imaginary part
of the integrand evaluated just above the real axis. Thus, we have
\beq\label{fg-contour-Im} 0 = I = \oint_{\Gamma} {G_1(z)
G_2(z)\over z-x}\,{dz\over 2\pi i} = -{1\over\pi}\,\Im\,
\int\limits_{-\infty}^\infty {G_1(y+i0)G_2(y+i0)\over y-x +i0}\,
dy\,. \eeq The last integrand may be written as
$$\pi^2\left({P\over y-x} -i\pi\delta(y-x)\right)\prod_{k=1,2}\left(\bar\rho_k^H(y) +
i\bar\rho_k(y)\right)\,,$$ by virtue of (\ref{PS}) and (\ref{G2}).
Upon substituting the last expression in (\ref{fg-contour-Im}) and
taking the imaginary part, we obtain the desired result
(\ref{fgHilbert1}). Note that for $\rho_1=\rho_2 = \rho$,
(\ref{fgHilbert}) simplifies into \beq\label{ffHilbert}
2(\rho\rho^H)^H = (\rho^H)^2 - \rho^2 + \rho_0^2\,. \eeq

Finally, we shall prove an identity involving three functions
$\rho_k(x)\,\, (k=1,2,3)$ and their Hilbert transforms. Our proof
follows essentially the one given in \cite{jev-sak-PRD22, onofri}
for the case $\rho_1=\rho_2=\rho_3$, which is reproduced also in
the text-book \cite{sakita}. Let $G_k(z)$ be the resolvent
corresponding to $\bar\rho_k (x)$. Consider now the contour
integral \beq\label{fgh-contour} J = \oint_{\cal C_\infty}
\,{dz\over 2\pi i}\,G_1(z) G_2(z) G_3(z)\,.\eeq As in the previous
proof, due to the asymptotic behavior (\ref{Gasympt}) of the
resolvents, evidently \beq\label{fgh-contour-null} J = 0\,.\eeq
Since the resolvents are analytic off the real axis, we can deform
${\cal C_\infty}$ into the contour $\Gamma$, as in the previous
proof, which picks up the imaginary part of the integrand
evaluated just above the real axis. Thus, we have
\beq\label{fgh-contour-Im} 0 = J = -{1\over\pi}\,\Im\,
\int\limits_{-\infty}^\infty G_1(y+i0)G_2(y+i0)G_3(y+i0)\, dy\,.
\eeq The last integrand may be written as
$$-\pi^3\prod_{k=1}^3\left(\bar\rho_k^H(y) +
i\bar\rho_k(y)\right)\,,$$ by virtue of (\ref{G2}). Upon
substituting the last expression in (\ref{fgh-contour-Im}) and
taking the imaginary part, we obtain the desired result
\beq\label{cubic-id}
\int\limits_{-\infty}^\infty\,\left(\bar\rho^H_1\bar\rho^H_2\bar\rho_3
+ \bar\rho_1^H\bar\rho_2\bar\rho^H_3 +
\bar\rho_1\bar\rho^H_2\bar\rho^H_3 \right)\, dx =
\int\limits_{-\infty}^\infty\,\bar\rho_1\bar\rho_2\bar\rho_3\,dx\,.\eeq
Note that for $\bar\rho_1=\bar\rho_2 = \bar\rho_3 = \bar\rho$,
(\ref{cubic-id}) simplifies into \beq\label{fff-cuibic} 3
\int\limits_{-\infty}^\infty\,\bar\rho (\bar\rho^H)^2\, dx =
\int\limits_{-\infty}^\infty\,(\bar\rho)^3\, dx\,,\eeq which is
the identity proved in \cite{sakita, jev-sak-PRD22, onofri}.

Since (\ref{cubic-id}) holds for any triplet of functions
$\bar\rho_k$ in the class of functions thus considered, we can
write it formally as an identity among distributions acting upon
these test functions, namely, the well-known \cite{jackiw}
identity \beq\label{PPdelta} {P\over x-y}{P\over x-z} + {P\over
y-x}{P\over y-z} + {P\over z-x}{P\over z-y} =
\pi^2\delta(x-y)\delta(x-z)\,.\eeq In \cite{jackiw}, the identity
(\ref{PPdelta}) was proved using Fourier transforms.
For alternative proofs of the identities discussed in this Appendix, and 
for more information about Hilbert-transform techniques, 
see Appendix A of \cite{abanov1}.

As should be clear from our proof, (\ref{PPdelta}) holds only when
the distributions on both its sides act upon functions which are
integrable on the whole real line. However, this identity is
frequently used in the literature on collective field theory
beyond its formal domain of validity. For further discussion of
this problem see Appendix B, where we show that this
transgression is benign, and can be compensated for by readjusting
the chemical potential which governs the normalization condition
(\ref{conservation}).

\pagebreak

%


\newpage
\setcounter{equation}{0}
\setcounter{section}{0}
\renewcommand{\theequation}{B.\arabic{equation}}
\renewcommand{\thesection}{Appendix B:}
\section{A Paradox and its Resolution}
\vskip 5mm
\setcounter{section}{0}
\renewcommand{\thesection}{B}

The expression for the collective potential in (\ref{Vcoll})
contains bilocal as well as trilocal terms in the density. It is
customary in the literature to avoid the trilocal terms by
applying a standard procedure as follows: The principal value
distribution, acting on functions integrable along the whole real
line, satisfies the identity (\ref{PPdelta}), which we rewrite
here for convenience
\begin{equation}
\frac{P}{x-y} \frac{P}{x-z} + \frac{P}{y-z} \frac{P}{y-x} +
\frac{P}{z-x} \frac{P}{z-y} =   {\pi}^{2} \delta(x-y)
\delta(x-z)\,. \label{principal}
\end{equation}
Making use of (\ref{principal}) in (\ref{Vcoll}), we obtain
\beqra\label{Vcoll1} \tilde V_{coll} &=& \frac{(\lambda\pi)^{2}}{6
m} \int\, dx\, \rho^3
 + \frac{{(\lambda - 1)}^{2}}{8 m} \int\, dx\, \frac{{(\partial_{x}
\rho)}^{2}}{\rho} + \frac{ \lambda (\lambda - 1)}{2 m} \int\, dx
\,\partial_{x} \rho \; \pv \int dy
\frac{\rho(y)}{x-y}\nonumber\\{}\nonumber\\ &+& \tilde\mu \left(N
- \int\, dx\, \rho(x) \right) \eeqra This expression for $\tilde
V_{coll}$ is evidently devoid of any trilocal terms. (Note that
the chemical potential $\tilde \mu$ in (\ref{Vcoll1}) need not
coincide with the one in (\ref{Vcoll}), as our notations imply.)

The classical equation of motion which results from varying
(\ref{Vcoll1}) is \beq\label{variationaleq1}
\frac{(\lambda\pi)^{2}}{2 m} \rho^2 - \frac{{(\lambda - 1)}^{2}}{8
m} { ( \frac{\partial_{x} \rho} {\rho})}^{2} - \frac{{(\lambda -
1)}^{2}}{4 m}\partial_{x} ( \frac{\partial_{x} \rho}{\rho}) -
\frac{\lambda (\lambda - 1)}{m} \pv \int dy \frac{\partial_{y}
\rho(y)}{x-y} = \tilde\mu\,. \eeq It was this form of the equation
of motion (rather than (\ref{variationaleq})), from which the
solitons and density waves were derived in the pioneering
work \cite{Polychronakos:1994xg}.

It can be checked that $\rho_s$ in (\ref{rhos}), $\rho_p$ in
(\ref{LAW}) and $\rho_n$ in (\ref{negLAW}), the solutions of the
variational equation (\ref{variationaleq}) of the first form
(\ref{Vcoll}) of the collective potential, are also solutions of
(\ref{variationaleq1}) (albeit, with values of $\tilde\mu$
different from those of (\ref{mu})). That this is true may look
surprising, and even paradoxical to some readers, since neither of
these solutions is integrable along the whole real line, which is
a necessary condition for (\ref{principal}) to hold. This should
be clear from the proof of (\ref{PPdelta}) in Appendix A, but it
can also be demonstrated by a simple counter example - just apply
both sides of (\ref{principal}) on three constant functions and
integrate over all coordinates. The LHS would be null, while the
RHS would diverge.

In fact, the latter counter example is precisely relevant to
determining the ground state of the collective Hamiltonian
(\ref{Hcollective}). The uniform ground state density
$\rho=\varrho_0$ is a solution of the BPS equation (\ref{BPS}),
and of course, also of the variational equation
(\ref{variationaleq}) with $\mu=0$. The energy density tied in it
is of course null. It is also a solution of the alternative
variational equation (\ref{variationaleq1}) with $\tilde\mu =
{(\lambda\pi\varrho_0)^2\over 2m}$ and energy density (with
respect to (\ref{Vcoll1})) ${(\lambda\pi)^2\varrho_0^3\over 6m}$.

Thus, it seems that using (\ref{principal}) beyond its formal
domain of validity is a mild transgression, which is compensated
for by appropriately readjusting the chemical potential. This is
indeed true, as we shall now prove, thus resolving the paradox why
(\ref{variationaleq}) and (\ref{variationaleq1}) always lead to
the same solutions. To this end we shall consider all $\rho$
configurations which are simultaneous solutions of
(\ref{variationaleq}) and (\ref{variationaleq1}). Such functions
are evidently extrema of $\Delta V = V_{coll} - \tilde V_{coll}$.
From (\ref{Vcoll}) and (\ref{Vcoll1}) we obtain
\beqra\label{deltaV} \Delta V  &=& {(\lambda\pi)^2\over 6m}
\,\Im\,\int\,dx\,(\rho^H + i\rho)^3  -
(\mu-\tilde\mu)\,\int\, dx\,\rho\nonumber\\{}\nonumber\\
&=& {(\lambda\pi)^2\over 6m} \,\Im\,\int\,dx\,\Phi^3(x +i0)
-(\mu-\tilde\mu)\,\Im\,\int\,dx\,\Phi(x +i0)\,.\eeqra (Note
that we have omitted from this expression the constant term
$(\mu-\tilde\mu)N\,.$) Due to the analytic structure of $\Phi
(z)$, and as explained in Appendix A, the latter integral can be
written as the contour integral \beq\label{deltaV1} \Delta V =
-{\lambda^2\pi^3\over 6m} \,\oint_{{\cal C}_\infty}\,{dz \over 2\pi
i}\, \Phi^3(z) + \pi(\mu-\tilde\mu)\,\oint_{{\cal C}_\infty}\,{dz
\over 2\pi i}\, \Phi(z) \,,\eeq where ${\cal C}_\infty$ is a
circle of infinite radius, centered at the origin. Note that $\Phi
(z)$ need not decay as $z\rightarrow\infty$, since $\int\, dx\,
\rho$ may diverge. Thus, in general $\Delta V\neq 0$.

We shall now determine solutions of \beq\label{deltadeltaV0}
{\delta \Delta V\over \delta \rho(x)} = 0 \,.\eeq To this end, let
us compute \beq\label{deltaPhi} {\delta \Phi (z)\over \delta
\rho(x)} = -{1\over\pi}{\delta\over\delta
\rho(x)}\,\int\limits_{-\infty}^\infty\,{\rho(u)\,du\over z-u} =
-{1\over\pi}\, {1\over z-x}\,.\eeq From this we infer that
\beq\label{deltadeltaV} {\delta \Delta V\over \delta \rho(x)} =
{(\lambda\pi)^2\over 2m}\,\oint_{{\cal C}_\infty}\,{dz \over 2\pi
i}\, {\Phi^2(z)\over z-x} - (\mu-\tilde\mu)\,.\eeq The contour
${\cal C}_\infty$ in the last integral can be deformed to the
countour $\Gamma$, defined in Appendix A, which essentially
picks up the imaginary part of the integrand evaluated just above
the real axis. Thus, in a manner similar to the discussion in 
Appendix A, from (\ref{fgHilbert}) to (\ref{ffHilbert}), we obtain
\beq\label{deltadeltaV1} {\delta \Delta V\over \delta \rho(x)} =
{(\lambda\pi)^2\over 2m}\,\left[ (\rho^H)^2 - \rho^2 -
(2\rho\rho^H)^H\right] - (\mu-\tilde\mu)\,.\eeq But from the
identity (\ref{ffHilbert}) we see that the latter equation boils
down to \beq\label{deltadeltaV2} {\delta \Delta V\over \delta
\rho(x)} = \tilde\mu -\mu - {(\lambda\pi\rho_0)^2\over 2m}\,.\eeq
In other words, the condition (\ref{deltadeltaV0}) simply relates
the two chemical potentials \beq\label{deltamu} \tilde\mu = \mu +
{(\lambda\pi\rho_0)^2\over 2m}\,,\eeq setting no further
conditions on $\rho (x)$, where $\rho_0$ is the subtraction
constant associated with the $\rho$ in question, and should not be
confused with the one appearing in (\ref{rhos}).

To summarize - any solution of (\ref{variationaleq}) with chemical
potential $\mu$ is simultaneously a solutions of
(\ref{variationaleq1}) with chemical potential $\tilde\mu$ given
by (\ref{deltamu}).

\pagebreak

%


\newpage
\setcounter{equation}{0}
\setcounter{section}{0}
\renewcommand{\theequation}{C.\arabic{equation}}
\renewcommand{\thesection}{Appendix C:}
\section{A Brief Summary of the Collective Field Formulation of the Calogero Model}
\vskip 5mm
\setcounter{section}{0}
\renewcommand{\thesection}{C}

In order for this paper to be self-contained, we briefly 
summarize in this appendix the derivation of the collective-field Hamiltonian (\ref{Hcollective})  from (\ref{Hcalogero}) .

The singularities of the Calogero-model Hamiltonian (\ref{Hcalogero}), namely, 
\begin{equation} \label{h1C}
H = - \frac{1}{2 m} \sum_{i=1}^{N} 
\frac{{\partial}^{2}}{\partial {x_{i}}^{2}} +
\frac{\lambda (\lambda - 1)}{2 m} \sum_{i \neq j
}^{N} \frac{1}{{(x_{i} - x_{j})}^{2}}\,,
\end{equation}
at points where particles coincide, implies that the many-body eigenfunctions 
contain a Jastrow-type prefactor 
\beq\label{jastrowC}
\Pi = \prod_{i < j}^{N} { (x_{i} - x_{j})}^{\lambda}\,.
\eeq
This Jastrow factor vanishes (for positive $\lambda$) at particle coincidence points, and 
multiplies that part of the wave-function which is totally symmetric under any permutation
of particles\footnote{Note, in particular, that for $ \; \lambda=0 \; $ and 
$ \;\lambda =1, \; $ the model describes interacting 
bosons and fermions, respectively.}. It is precisely these symmetric wave-functions
on which the collective field operators act, as explained below. 

Let us recall at this point some of the basic ideas of the 
collective-field method \cite{sakita,Jevicki:1979mb,JFnmm}, adapted specifically to the Calogero 
model\cite {AJL,Andric:1994su}: 
Instead of solving the Schr\"odinger equation associated with (\ref{h1C}) 
for the many-body eigenfunctions, subjected to the appropriate particle 
statistics (Bosonic, Fermionic of fractional), we restrict ourselves to 
functions which are totally symmetric under any permutation of identical 
particles. This we achieve by stripping off the Jastrow factor 
(\ref{jastrowC}) from the eigenfunctions, which means performing on 
(\ref{h1C}) the similarity transformation
\begin{equation} \label{similaritytrC}
 H \rightarrow \tilde H = \Pi^{-1} H \Pi \,,
\end{equation}
where the Hamiltonian
\begin{equation} \label{h1trC}
 \tilde H = - \frac{1}{2 m} \sum_{i=1}^N \frac{{\partial}^{2}}
{\partial {x_{i}}^{2}} -
    \frac{\lambda}{m} \sum_{i \neq j}^N
   \frac{1}{x_{i} - x_{j}}\,\frac{\partial}{\partial x_i} \,.
   \end{equation}
Note that $\tilde H$ does not contain the singular two-body interactions. 
By construction, this Hamiltonian is hermitian with respect to the measure 
$$d\mu (x_i) = \Pi^2\, d^N x\,,$$ 
(as opposed to the original Hamiltonian $H$ in (\ref{h1C}), which is 
hermitian with respect to the flat Cartesian measure).

We can think of the symmetric many-body wave-functions acted upon by 
$\tilde H$ as functions depending on all possible symmetric
combinations of particle coordinates. These combinations form an overcomplete set of 
variables. However, as explained below, in the {\em continuum} limit, 
redundancy of these symmetric variables has a negligible effect. 
The set of these symmetric variables can be generated, for example, 
by producs of moments of the collective  - or density - field
\beq\label{collectiveC}
\rho (x) = \sum_{i = 1}^N \delta( x - x_{i})\,.
\eeq
The collective-field theory for the Calogero model is obtained
by changing variables from the particle coordinates $ \; x_{i} \; $ 
to the density field $ \; \rho (x) \; $. This transformation replaces the finitely many variables
$ \; x_{i} \; $ by a continuous field, which is 
just another manifestation of overcompleteness of the collective variables.  
Clearly, description of the particle systems in terms of continuous fields
becomes an effectively good description in the high density limit.  Of course, the large 
density limit  means that we have taken the large- $N$ limit.

Changing variables from particle coordinates $x_i$ to the 
collective fields (\ref{collectiveC}) implies that we should express all 
partial derivatives in  the Hamiltonian $\tilde H$ in (\ref{h1trC}) as 
\beq\label{derivativesC} 
\frac{\partial}{\partial x_{i}} = \int dx \frac{\partial \rho(x)}
{\partial x_{i}} \frac{\delta}{\delta \rho(x)}\,,\end{equation}
where we applied the differentiation chain rule. 

In the large $- N$ limit, the Hamiltonian $\tilde H$ can be 
expressed entirely in terms of the collective field $ \; \rho (x)$ and its canonical conjugate momentum 
\beq\label{momentaC}
\pi(x) = - i \frac{\delta}{\delta \rho(x)}\,,
\end{equation}
as we show below. It follows from (\ref{derivativesC}) and (\ref{momentaC}) 
that the particle momentum operators (acting on symmetric wave-functions)
may be expressed in terms of the collective-field momenta at particular
points on the line as 
\beq\label{particlemomentaC}
p_i = -\pi'(x_i)
\eeq
(where $\pi'(x) = \pax \pi (x)$). 
Finally, note from (\ref{collectiveC}) that the collective field obeys the 
normalization condition
\beq\label{conservationC}
\int dx \rho(x) = N\,.
\eeq
The density field $ \; \rho\; $ and its conjugate 
momentum $ \; \pi\; $ satisfy the equal-time canonical 
commutation relations\footnote{According to 
(\ref{conservationC}), the zero-momentum modes of the density fields are 
constrained, i.e., non-dynamical. This affects the commutation 
relation (\ref{canonicalC}), whose precise form is $[\rho(x), \pi (y)]
= i (\delta(x - y) - (1/l))$, where $l$ is the size of the large 
one-dimensional box in which the system is quantized, which is much larger 
than the macroscopic size $L$ of the particle condensate in the system. 
In what follows, we can safely ignore this $1/l$ correction in the commutation
relations.}
\beq\label{canonicalC}
[\rho (x), \pi (y)] = i \delta(x - y)\,,
\eeq
(and of course $[\rho (x), \rho (y)] = [\pi (x), \pi (y)] = 0)\,. $
By substituting (\ref{collectiveC})-(\ref{particlemomentaC}) in (\ref{h1trC}), 
we obtain the continuum-limit expression for $\tilde H$ as 
\begin{equation}\label{htildeC}
\tilde H = \frac{1}{2 m} \int dx \rho(x) { ( \partial_{x} 
\pi(x))}^{2}
 - \frac{i}{m} \int dx \rho(x) \left( \frac{\lambda - 1}{2} 
\frac{\partial_{x} \rho}{\rho} +
\lambda \pv \int \frac{ dy \rho(y)}{x - y}  \right) \partial_{x} \pi(x)
\end{equation}
where $ \; \pv \int \; $ denotes Cauchy's principal value.

It can be shown \cite{sakita} that (\ref{htildeC}) is hermitian with 
respect to the functional measure\footnote{By definition (recall 
(\ref{collectiveC})), this measure is defined only over positive values of 
$\rho$\,.}
\beq\label{functionalmeasureC} 
{\cal D} \mu [\rho] = J[\rho] \prod_x 
d\rho(x)\,,
\eeq
where $ J[\rho] $ is the Jacobian of the transformation from the 
$ \; \{ x_{i} \} \; $ to the collective field 
$ \; \{ \rho(x) \}$\,. In the large - $N$ limit 
it is given by \cite{Bardek:2005yx} 
\begin{equation}
 \ln J = (1 - {\lambda}) \int dx \rho(x) \ln \rho(x) 
 - {\lambda} \int dx dy \rho(x) \ln |x - y | \rho(y)
\end{equation}
It is more convenient to work with a Hamiltonian, which unlike (\ref{htildeC}),
is hermitian with respect to the flat functional Cartesian measure
$\prod_x d\rho(x)\,.$ This we achieve by means of the 
similarity transformation $\psi \rightarrow J^{\frac{1}{2}} \psi\,, 
\tilde H \rightarrow H_{coll} = J^{\frac{1}{2}} \tilde H J^{- \frac{1}{2}}\,,$
where the continuum {\em collective} Hamiltonian is 
\begin{equation} \label{colhamC}
 H_{coll}  = \frac{1}{2 m} \int dx\, \pi'(x)\, \rho(x)\, \pi'(x) +
  \frac{1}{2 m} \int dx \rho(x) {\left( \frac{\lambda - 1}{2} 
\frac{\partial_{x} \rho}{\rho} +
\lambda \pv \int \frac{ dy \rho(y)}{x - y}   \right)}^{2}
 +  H_{sing},~~~~~~~~~~~~~~~~~~~~~~~~~~~~~~~~~~~~~~~~~~~~~~~~~~~~~~~~~~~~~~~~~~~~~~~~~~~~~~~~~~~~~~~~~~~~~~~~~~~~~~~~~~~~~~~~~~~~~~~~~~~~~~~~~
\end{equation}
namely, the Hamiltonian given by (\ref {Hcollective}) and (\ref{Hsing}). 
The collective-field Hamiltonian (\ref{Hcollective2F}) of the two-family Calogero model can be derived from (\ref{h1}) in a similar manner.


\pagebreak

{\bf Acknowledgement}\\
This work was supported in part by the Ministry of Science and Technology of the Republic of Croatia 
under contract No. 098-0000000-2865 and by the US National Science Foundation  
under Grant No. PHY05-51164.  


\end{document}